\documentstyle[prl,aps,multicol,psfig]{revtex}
\begin{document}

\draft

\title{Superfluidity of spin-polarized ${\bf ^6}$Li}

\author{H.T.C. Stoof,$^1$ M. Houbiers,$^1$
        C.A. Sackett,$^2$ and R.G. Hulet$^2$}
\address{$^1$Institute for Theoretical Physics,
             University of Utrecht, Princetonplein 5, \\
             P.O. Box 80.006, 3508 TA  Utrecht,
             The Netherlands \\
         $^2$Physics Department and Rice Quantum Institute,
             Rice University, \\
             P.O. Box 1892, Houston, Texas 77251}

\maketitle

\begin{abstract}
We study the prospects for observing superfluidity in a spin-polarized atomic
gas of $^6$Li atoms, using  state-of-the-art interatomic potentials. We
determine the spinodal line and show that a BCS transition to the superfluid
state can indeed occur in the (meta)stable region of the phase diagram if the
densities are sufficiently low. Moreover, for a total density of
$10^{12}~cm^{-3}$, which still fulfills this requirement, we find a critical
temperature of only $29~nK$. We also discuss the stability of the gas due to
exchange and dipolar relaxation and conclude that the prospects for observing
superfluidity in a magnetically trapped atomic $^6$Li gas are particularly
promising for magnetic bias fields larger than $10~T$.
\end{abstract}


\begin{multicols}{2}

Ultracold atomic gases have received much attention in recent years, because of
their novel properties. For instance, these gases are well suited for
high-precision measurements of single-atom properties and for the observation
of collisional and optical phenomena that reflect the (Bose or Fermi)
statistics of the constituent particles. Moreover, a large variety of
experimental techniques are available to manipulate the atomic gas samples by
means of electromagnetic fields \cite{chu}, which offers the exciting
possibility to achieve the required conditions for quantum degeneracy and to
study macroscopic quantum effects in their purest form.

At present, most experimental attempts towards quantum degeneracy have been
performed with bosonic gases and have been aimed at the achievement of
Bose-Einstein condensation. In particular, most of the earlier experiments used
atomic hydrogen \cite{greytak,walraven}. These experiments provided crucial
ingredients for the recent attempts with alkali vapors, for which the
experimental advances towards the degeneracy regime were so rapid that
Bose-Einstein condensation has actually been reported now for the isotopes
$^{87}$Rb \cite{eric} and $^7$Li \cite{randy}.

In view of these exciting developments it seems timely to investigate
theoretically also the properties of spin-polarized atomic $^6$Li, since $^6$Li
is a stable fermionic isotope of lithium that can be trapped and cooled in much
the same way as its bosonic counterpart. Therefore, magnetically trapped $^6$Li
promises to be an ideal system to study degeneracy effects in a
weakly-interacting Fermi gas, thus providing valuable complementary information
on the workings of quantum mechanics at the macroscopic level. Moreover, using
a combination of theoretical \cite{ard-jan,dalgarno} and experimental
\cite{abraham} results, accurate knowledge of the interparticle (singlet and
triplet) potential curves of lithium have recently been obtained which lead to
the prediction of a large and negative s-wave scattering length $a$ of $-4.6
\cdot 10^3~a_0$ ($a_0$ is the Bohr radius) for a spin-polarized $^6$Li gas.

This is important for two reasons: First, the fact that the scattering length
is negative implies that at the low temperatures of interest ($\Lambda \gg
r_V$, where
$\Lambda = (2\pi\hbar^2/mk_BT)^{1/2}$ is the thermal de Broglie wavelength of
the atoms and $r_V$ is the range of the interaction) the effective interaction
between the lithium atoms is attractive and we expect a BCS-like phase
transition to a superfluid state at a critical temperature
\begin{equation}
\label{tc}
T_c \simeq \frac{5 \epsilon_F}{3 k_B}
             \exp \left\{ -\frac{\pi}{2k_F|a|} - 1 \right\}~,
\end{equation}
with $\epsilon_F = \hbar^2 k_F^2/2m$ the Fermi energy of the gas \cite{ef}.
Secondly, we see that the critical temperature depends exponentially on
$1/k_F|a|$ which usually, when the magnitude of $a$ is of the order of the
range of the interaction $r_V$, is very large in the dilute limit $k_Fr_V \ll
1$. Therefore, it was previously concluded that the BCS transition in a dilute
fermionic system (in particular spin-polarized deuterium) is experimentally
unattainable \cite{tony}. However, with the anomalously large scattering length
of $^6$Li this conclusion needs revision as we will see now in more detail.

Since $^6$Li has an electron spin of $s=1/2$ and a nuclear spin of $i=1$, the
$1s$ groundstate of $^6$Li consists of six hyperfine levels which are labeled
by $|1\rangle$ through $|6\rangle$ in such a way that their energy increases at
small magnetic fields. At zero magnetic field these correspond exactly to the
states $|f,m_f\rangle$ with a total spin $f$ of either $1/2$ or $3/2$ and a
hyperfine splitting of $3a_{hf}/2 \simeq 10.95~mK$. In a conventional trapping
experiment one tries to achieve both electron and nuclear spin-polarization by
trapping only atoms in the doubly spin-polarized state $|6\rangle \equiv
|m_s=1/2,m_i=1\rangle$. In the case of $^6$Li, however, such a procedure
creates a gas in which the atoms only interact extremely weakly because the
Pauli principle now forbids s-wave scattering. Therefore, we cannot take
advantage of the large negative value of the triplet scattering length and the
BCS-transition temperature is unattainable.

To avoid this problem we need to trap two hyperfine states. This can be
achieved most easily by using relatively large magnetic fields $B \gg
a_{hf}/\mu_e \simeq 0.011 T$ ($\mu_e$ is the electron magnetron), because then
the electron and nuclear spins are almost decoupled and the states
$|4\rangle \simeq |m_s=1/2,m_i=-1\rangle$,
$|5\rangle \simeq |m_s=1/2,m_i=0\rangle$, and $|6\rangle$ can all be trapped.
Moreover, the trap can, for example, be loaded by creating first a doubly
spin-polarized gas in the conventional manner and subsequently applying a
microwave pulse to populate one of the other trapping states. Notice that
because the gas is now only electron (and not nuclear) spin polarized, there is
a large cross-section $4\pi a^2$ for the thermalizing collisions required for
evaporative cooling. In addition, this implies that the gas is in thermal
equilibrium in the spatial degrees of freedom, even though the spin degrees of
freedom are not. Notice also that the electron spin polarization is not
complete because the states $|4\rangle$ and $|5\rangle$ have a small admixture
(of $O(a_{hf}/\mu_eB)$) of $|m_s=-1/2,m_i=0\rangle$ and
$|m_s=-1/2,m_i=1\rangle$, respectively. Therefore, two atoms do not interact
solely via the triplet interaction. When we consider the lifetime of the gas,
however, we find that the influence of this can be neglected if the magnetic
field is larger than $1~T$.

Although it is possible to study any combination of the three trapping states,
we will consider here only a gas in which the atoms are in a mixture of state
$|5\rangle$ or state $|6\rangle$ because this minimizes the number of decay
processes. Furthermore, we analyze here first the homogeneous case. The
influence of the trapping potential will be discussed in a separate
publication. Taking only s-wave scattering into account and following Ref.\
\cite{henk} to include all two-body processes, we can determine the
thermodynamic properties of this gas by considering the hamiltonian
\begin{eqnarray}
H = \int d\vec{x}~ \left\{ \sum_{\alpha=5}^6
      \psi_{\alpha}^{\dagger}(\vec{x})
      \left(- \frac{\hbar^2 \nabla^2}{2m} - \mu'_{\alpha} \right)
      \psi_{\alpha}(\vec{x}) \right.                 \nonumber \\
   + \Delta_0 \psi_6^{\dagger}(\vec{x})\psi_5^{\dagger}(\vec{x})
   + \Delta_0^* \psi_5(\vec{x})\psi_6(\vec{x})       \nonumber \\
   \left. - \frac{|\Delta_0|^2}{V_0}
          - \frac{4\pi a\hbar^2}{m}~n_5 n_6 \right\}~.
                                               \hspace*{-0.15in}
\end{eqnarray}
Here the field $\psi_{\alpha}^{\dagger}(\vec{x})$ ($\psi_{\alpha}(\vec{x})$)
creates (annihilates) an atom at position $\vec{x}$ in a spin state
$|\alpha\rangle$ that has a (renormalized) chemical potential $\mu'_{\alpha}$
and an average density $n_{\alpha}$. In addition,
$V_0 = \int d\vec{x}~V_T(\vec{x})$ is the zero momentum component of the
triplet potential and
$\Delta_0 = V_0 \langle \psi_5(\vec{x})\psi_6(\vec{x}) \rangle$
is the equilibrium value of the appropriate BCS order parameter.

A first condition for the observability of the BCS transition is that it takes
place in the (meta)stable region of the phase diagram \cite{henk}. The
requirements on the densities in the two hyperfine levels that result from this
condition of mechanical stability of the gas can be obtained as follows. For
the normal state (where $\Delta_0 = 0$) the hamiltonian is diagonal and we can
immediately compute the pressure $p$ of the gas by evaluating the thermodynamic
potential density $\Omega/V = -p$. This results in
\begin{eqnarray}
p = k_BT \sum_{\alpha}
         \int \frac{d\vec{k}}{(2\pi)^3}~
              ln \left(1 +
                 e^{-\beta(\epsilon(\vec{k})-\mu'_{\alpha})}
                 \right)                           \nonumber \\
      + \frac{4\pi a\hbar^2}{m}~n_5 n_6~, \hspace*{-0.15in}
\end{eqnarray}
where $\epsilon(\vec{k}) = \hbar^2 \vec{k}^2/2m$ is the kinetic energy of the
atoms and $\beta = 1/k_BT$. Furthermore, the density of lithium atoms in the
two hyperfine states is determined by
\begin{equation}
\label{n}
n_{\alpha} = \int \frac{d\vec{k}}{(2\pi)^3}~
                                   N_{\alpha}(\vec{k})~,
\end{equation}
introducing the notation $N_{\alpha}(\vec{k})$ for the occupation numbers which
in our case are equal to the Fermi distribution function $(e^{\beta x}+1)^{-1}$
evaluated at $\epsilon(\vec{k})-\mu'_{\alpha}$.

\begin{figure}
\psfig{figure=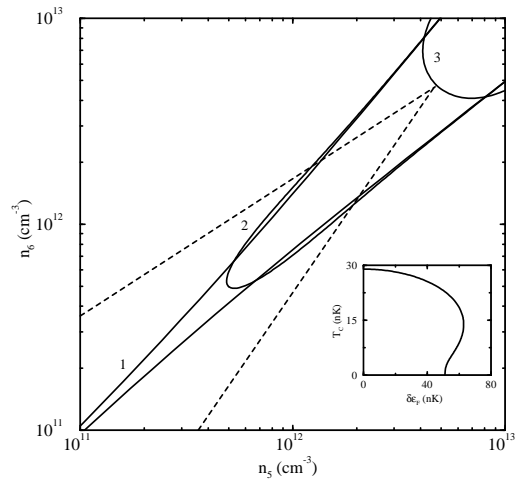}
\caption{\narrowtext
         The required conditions for the achievement of the
         BCS transition in the ($n_5$, $n_6$) plane.
         The dashed curve shows the spinodal line,
         and curves 1, 2 and 3 give the
         conditions at which the critical temperature is $0$,
         $29$ and $400~nK$, respectively. The inset shows the
         critical temperature as a function of $\delta\epsilon_F$
         for $\epsilon_F = 385~nK$.
         \label{fig1}}
\end{figure}

In the degenerate regime, neglecting corrections of $O((k_BT/\epsilon_F)^2)$,
the equation of state Eq.\ (\ref{n}) can be easily inverted and we find for the
pressure
\begin{equation}
p = \sum_{\alpha} n_{\alpha}k_BT~ \frac{1}{5}
          \left( 3 \sqrt{\frac{\pi}{2}}
                 n_{\alpha} \Lambda^3 \right)^{2/3}
    + \frac{4\pi a\hbar^2}{m}~n_5 n_6~.
\end{equation}
For the mechanical stability of the gas we must require that
$\partial p/\partial n_{\alpha} \ge 0$. Using the above result this leads to
two conditions on the densities in the two hyperfine levels, namely
\begin{eqnarray}
n_5 \le \frac{1}{6|a|}
          \left( 3n_6\sqrt{\frac{\pi}{2}} \right)^{2/3}
{\rm ~and~}
n_6 \le \frac{1}{6|a|}
          \left( 3n_5\sqrt{\frac{\pi}{2}} \right)^{2/3},
                                                   \nonumber
\end{eqnarray}
which have to be fulfilled simultaneously. The line where one of the equalities
holds is called the spinodal line and it is shown for $^6$Li in Fig.\
\ref{fig1}. Note that for the highest metastable densities we have $k_F|a| =
\pi/2$. Therefore, the ratio $k_BT_c/\epsilon_F$ is at most $0.23$ and our
determination of the spinodal line is self consistent for the temperatures of
interest.

Within this density region (i.e.\ also for $n_5 \neq n_6$) we can now consider
the critical temperature of the gas. For that we need to derive the BCS gap
equation. This can be achieved most easily by diagonalizing the hamiltonian by
means of a Bogoliubov transformation and then calculating the equilibrium value
of
$\Delta_0 = V_0 \langle \psi_5(\vec{x})\psi_6(\vec{x}) \rangle$.
In the limit of vanishing $\Delta_0$, or equivalently
$T \rightarrow T_c$, this procedure leads to the linearized BCS gap equation
\begin{equation}
\label{gap1}
\frac{1}{V_0} + \int \frac{d\vec{k}}{(2\pi)^3}~
    \frac{1 - N_5(\vec{k}) - N_6(\vec{k})}
         {2(\epsilon(\vec{k}) - \epsilon_F)} = 0~,
\end{equation}
where we have introduced the appropriate Fermi energy
$\epsilon_F = (\mu'_5 + \mu'_6)/2$. As expected, the integral in the left-hand
side of Eq.\ (\ref{gap1}) has an ultraviolet divergence due to the neglect of
the momentum dependence of the triplet potential in the hamiltonian. However,
from the Lippmann-Schwinger equation for the two-body T-matrix \cite{glockle}
we find that this divergence is cancelled by a renormalization of $1/V_0$ to
$1/T^{2B}(\vec{0},\vec{0};0) = m/4\pi a\hbar^2$. Therefore, the critical
temperature is determined by the condition
\begin{equation}
\label{gap}
\frac{m}{4\pi a\hbar^2} +
 {\cal P} \int \frac{d\vec{k}}{(2\pi)^3}~
            \frac{N_5(\vec{k}) + N_6(\vec{k})}
              {2(\epsilon_F - \epsilon(\vec{k}))} = 0~,
\end{equation}
which is free of divergences because of the Fermi distibutions in the Cauchy
principle-value integral.

We have solved Eqs.\ (\ref{gap}) and (\ref{n}) numerically for a given density
in the two hyperfine states. Furthermore, we have verified that the energy
dependence of the triplet s-wave cross section is unimportant even though the
scattering length is large and negative. Physically the latter is a result of
the presence of an almost bound state near the continuum treshold and one might
therefore have expected a relatively strong energy dependence. Our final
results are summarized in Fig.\ \ref{fig1} where we have plotted the critical
values of $n_5$ and $n_6$ for three different temperatures. (They can, of
course, be calculated directly from Eq.\ (\ref{tc}) if $n_5=n_6$.) In
particular, we find that non-zero critical temperatures are only possible if
the `polarization' $|n_6-n_5|/(n_6+n_5)$ is less than  $3k_BT_c/2\epsilon_F$.
We also find for
$\delta\epsilon_F = |\mu'_6 - \mu'_5|$ around $2k_BT_c$ a reentrance behaviour,
because at a fixed value of $\epsilon_F$ the function $T_c(\delta\epsilon_F)$
is multivalued. This is shown more explicitly in the inset of Fig.\ \ref{fig1}
and can be proven analytically by showing that the linearized gap equation
reduces to
\begin{equation}
\frac{\delta\epsilon_F}{8 \epsilon_F} \simeq
     \left(1 + \frac{2\pi^2}{3}
              \left(\frac{k_BT_c}{\delta\epsilon_F}\right)^2
     \right)
       \exp\left\{ -\frac{\pi}{2k_F|a|} - 2 \right\}~,
\end{equation}
for $k_BT_c \ll \delta\epsilon_F \ll \epsilon_F$ and thus explaining the
square-root behavior of $T_c(\delta\epsilon_F)$ found in this regime.

Finally, we also need to consider the lifetime of spin-polarized $^6$Li. It is
well-known that spin-polarized atomic gases are not completely stable because
collisions between the atoms can induce transitions from one hyperfine level
$|\alpha\rangle$ to another. In the case of interest these transitions can be
caused both by the central (singlet and triplet) interactions $V^c(\vec{r})$
and by the magnetic dipole-dipole interactions $V^d(\vec{r})$. In general, the
zero temperature rate constant of a transition $\alpha\beta \rightarrow
\alpha'\beta'$ equals
\begin{equation}
\label{rate}
G = 2\pi^3\hbar^2mp_{\alpha'\beta'}
     \left| T_{\{\alpha'\beta'\}\ell'm',\{\alpha\beta\}00}
                               (p_{\alpha'\beta'},0) \right|^2~,
\end{equation}
with $p_{\alpha'\beta'}$ the momentum, and $\ell'$ and $m'$ the angular
momentum quantum numbers in the final state \cite{stoof}.

The central interaction $V^c(\vec{r})$ cannot flip the electron or nuclear
spin. Therefore, the only transition due to this interaction that is possible
in a gas consisting of atoms in the hyperfine states $|5\rangle$ and
$|6\rangle$ is
$65 \rightarrow 61$. The coupling matrix element between these states is
proportional to the exchange potential $V^{ex}(\vec{r})= V_T(\vec{r}) -
V_S(\vec{r})$, i.e.\ the difference between the triplet and singlet potentials.
More precisely, if $B \gg a_{hf}/\mu_e$ it is equal to $- (a_{hf}/\mu_eB)
V^{ex}(\vec{r})/2\sqrt{2}~$. Therefore, at sufficiently high magnetic fields
the coupling becomes small and the T-matrix element in Eq.\ (\ref{rate}) can be
calculated in the distorted-wave Born approximation leading to
\begin{equation}
G^{ex} \simeq
   \pi^3\hbar^2 a_{hf}^2
    \left( \frac{m}{2\mu_eB} \right)^{3/2}
    \left| \langle \Psi^{(-)}_{000}|V^{ex}
                            |\Psi^{(+)}_{001} \rangle \right|^2~,
\end{equation}
where $|\Psi^{(\pm)}_{\ell mS}\rangle$ are the exact scattering states of two
lithium atoms with angular momentum quantum numbers $\ell$ and $m$, total
electron spin $S$, and at the energies of the initial and final state,
respectively.

\begin{figure}
\psfig{figure=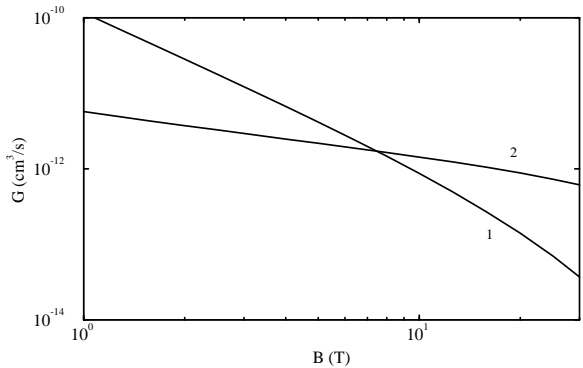}
\caption{\narrowtext
         The decay rate constants as a function of magnetic
         field. Curve 1 shows $G^{ex}$, whereas curve 2 gives
         $G^{1sf}$.
         \label{fig2}}
\end{figure}

The decay rate as a function of magnetic field following from this expression
is given in Fig.\ \ref{fig2}. We give the rate constant only for magnetic
fields larger than $1~T$ because we have verified numerically that only then
can the exchange interaction be treated in first order perturbation theory.
Note that the rate constant is much larger than the rate constant for the
similar process in magnetically trapped atomic hydrogen. This is again a result
of the fact that in the initial (triplet) state there is an almost bound state
near the continuum threshold which enormously enhances the amplitude of the
triplet wave function. Clearly, this is the drawback of having such a large
scattering length. Nevertheless, we see that a magnetic field of about $10~T$
suppresses the decay of the gas to the extent that for a density of
$10^{12}~cm^{-3}$, the lifetime is of the order of seconds.

Of the various magnetic dipole-dipole interactions the electron-electron
interaction is most important. Nevertheless, it is so weak that it can always
be treated by first-order perturbation theory \cite{stoof}. Thus, if $B \gg
a_{hf}/\mu_e$ we can neglect the hyperfine coupling between the electron and
nuclear spins and the total dipolar rate consist of the sum of a one spin-flip
and a two spin-flip contribution, i.e.\
$G^{d} = G^{1sf} + G^{2sf}$. Due to the different spin-matrix elements and the
difference in the energy released in the transition, we find the convenient
relation
$G^{2sf}(B) = 2G^{1sf}(2B)$. Hence, we need to consider only the one spin-flip
process, which we again treat in the distorted-wave Born approximation. For
magnetic fields larger than $1~T$ this leads to
\begin{equation}
G^{1sf} \simeq \frac{6}{5\pi} \sqrt{2m\mu_e B}~
               \frac{m(\mu_0\mu_e^2)^2}{\hbar^4}
               \left| r_{20} \right|^2~,
\end{equation}
where
$r_{20} = \int_0^{\infty} dr~ \psi^{(-)}_{21}(r) r^{-3}
                                    \psi^{(+)}_{01}(r)$ is the relevant
transition matrix element and the radial wavefunctions $\psi^{(\pm)}_{\ell
S}(r)$ are normalized as
\begin{equation}
\Psi^{(\pm)}_{\ell mS}(\vec{r}) \equiv
   \sqrt{\frac{2}{\pi\hbar^2}}
     \frac{\psi^{(\pm)}_{\ell S}(r)}{r}
                                  i^{\ell} Y_{\ell m}(\hat{r})~.
\end{equation}

The rate constant for the one spin-flip process is also shown in Fig.\
\ref{fig2}. Again, it is much larger than the rate constant of a similar proces
in atomic hydrogen. Notice that for magnetic fields larger than $10~T$ the
electron-electron dipolar rate dominates the decay and then also leads to a
lifetime of the order of seconds for a density of $10^{12}~cm^{-3}$. The decay
rates due to the electron-nucleus dipolar interaction are even smaller by a
factor of $(\mu_N/\mu_e)^2 \simeq 20 \cdot 10^{-6}$ and are completely
negligible. This proves that our assumption of a non-equilibrium distribution
in the spin-degrees of freedom is justified, because relaxation between the
hyperfine levels $|5\rangle$ and $|6\rangle$ requires a nuclear spin-flip and
will therefore take place on a timescale set by the electron-nucleus dipolar
rate which is much longer than the lifetime of the gas.

\vspace*{-0.5in}

In summary, we have shown that due to the large and negative triplet scattering
length the BCS transition to a superfluid state occurs at experimentally
accessible densities and temperatures in spin-polarized $^6$Li. It should be
pointed out however that the scattering length is extremely sensitive to the
interatomic potential, so the exact conditions required may be somewhat
different than that given here. A better estimate of the scattering length can
be obtained by repeating the experiment of Ref.\ \cite{abraham} for $^6$Li.
Moreover, as is well-known from liquid $^3$He, the phase below the critical
temperature is truly superfluid because it costs a finite amount of (free)
energy to have gradients in the phase of the order parameter $\Delta_0$. As a
result we have a macroscopic (free) energy barrier for the decay of superflow
and the gas can sustain persistent mass currents. By calculating the various
collisional decay rates in the gas, we have also shown that reasonable
lifetimes can be achieved even for densities as high as $10^{12}~cm^{-3}$ if
the bias magnetic field is larger than $10~T$. We hope that our work will
stimulate new experiments with this interesting quantum gas.

We are grateful to Eric Abraham for his help with the construction of the
interatomic potentials and to Michel Bijlsma for useful discussions. The work
at Rice is supported by the National Science Foundation and the Welch
Foundation.

\end{multicols}

\begin{references}
\bibitem{ef} This result follows from Eq.\ (\ref{gap}) below,
             in the case that the occupation numbers
             (i.e.\ $N_5(\vec{k})$ and $N_6(\vec{k})$) and
             therefore the densities and Fermi energies of the
             two spin states involved in the BCS transition are
             equal.
\end{references}
\end{document}